# Band-Like Transport in High Mobility Unencapsulated Single-Layer MoS$_2$ Transistors

Deep Jariwala[1], Vinod K. Sangwan[1], Dattatray J. Late[1,a)], James E. Johns[1], Vinayak P. Dravid[1], Tobin J. Marks[1,2], Lincoln J. Lauhon[1], and Mark C. Hersam[1,2,3,b)]

[1]Department of Materials Science and Engineering, Northwestern University, Evanston, Illinois 60208, USA.

[2]Department of Chemistry, Northwestern University, Evanston, Illinois 60208, USA.

[3]Department of Medicine, Northwestern University, Evanston, Illinois 60208, USA.

**Abstract**

Ultra-thin MoS$_2$ has recently emerged as a promising two-dimensional semiconductor for electronic and optoelectronic applications. Here, we report high mobility (>60 cm$^2$/Vs at room temperature) field-effect transistors that employ unencapsulated single-layer MoS$_2$ on oxidized Si wafers with a low level of extrinsic contamination. While charge transport in the sub-threshold regime is consistent with a variable range hopping model, monotonically decreasing field-effect mobility with increasing temperature suggests band-like transport in the linear regime. At temperatures below 100 K, temperature-independent mobility is limited by Coulomb scattering, whereas, at temperatures above 100 K, phonon-limited mobility decreases as a power law with increasing temperature.

[a)]Present address: National Chemical Laboratory, Pune, Maharashtra, India.

[b)]Author to whom correspondence should be addressed. Electronic mail: m-hersam@northwestern.edu.

**Manuscript**

The unique properties of two-dimensional (2D) graphene has led to growing interest in other 2D materials including the layered transition metal dichalcogenides such as molybdenum disulphide ($MoS_2$).[1,2] In contrast to gapless graphene, single-layer $MoS_2$ possesses a direct bandgap in addition to moderately high field-effect mobilities[3-8] and efficient light emission, which make it a promising candidate for low-power digital electronics[5,9] and optoelectronics.[10,11,12-14,15-17] Initial studies on unencapsulated single-layer $MoS_2$ field-effect transistors (FETs) reported mobilities of 0.2-12 $cm^2/Vs$[6,18-21] and temperature-dependent charge transport results that were consistent with variable range hopping[18] or thermally activated transport.[19] In contrast, thicker few-layer $MoS_2$ and encapsulated single-layer $MoS_2$ FETs have shown higher mobilities (60-500 $cm^2/Vs$)[3,4,19,22-24] and band-like transport,[7,19] suggesting that the charge transport mechanism in $MoS_2$ strongly depends on extrinsic sources of scattering (e.g., adsorbates).

In this Letter, we report high field-effect mobilities (>60 $cm^2/Vs$ at room temperature) in unencapsulated single layer $MoS_2$ FETs at high vacuum conditions (2 x $10^{-6}$ Torr). By comparing devices in and out of vacuum, it is apparent that atmospheric adsorbates strongly dope $MoS_2$ and degrade conductivity by more than an order of magnitude. In addition, we observe Mott variable range hopping (VRH) transport in the sub-threshold regime and band-like transport in the high carrier density linear regime, which suggests a reduced density of trap states at energies near the band edge. Variable temperature charge transport measurements also elucidate the principal scattering mechanisms as Coulomb scattering at low temperatures (< 100 K) and phonon scattering at high temperatures (> 100 K). Overall, these results provide insight into the factors that control charge transport in single-layer MoS$_2$, thus informing future efforts to realize high performance MoS$_2$ electronic and optoelectronic devices.

MoS$_2$ flakes were obtained *via* mechanical exfoliation using scotch tape on thermally oxidized (300 nm SiO$_2$) Si substrates. The MoS$_2$ flakes were directly exfoliated on solvent-cleaned substrates without subjecting them to any preprocessing (e.g., lithography, reactive ion etching, or plasma etching) in an effort to minimize contamination at the MoS$_2$/SiO$_2$ interface. Additional details on fabrication and materials are provided in Supplementary Material S1.[46] Single-layer flakes were identified using optical contrast microscopy and subsequently confirmed *via* Raman spectroscopy (Supplementary Material S2[46]).[25,26] Metal contact electrodes were patterned on the selected flakes by electron beam lithography and thermal evaporation of Au (without an adhesion layer of Ti or Cr). As previously reported, Au without any adhesion layer produces linear output characteristics (Figure 1(a)), suggesting electrically ohmic contacts.[6,18,27] Au contacts are also more resistant to degradation over time as compared to contacts with Ti or Cr adhesion layers (Supplementary Material S3[46]). The insets of Figures 1(a) and 1(b) provide a schematic and an optical micrograph of the resulting FET, respectively.

All the devices were measured in a vacuum probe station (Lakeshore Cryogenics) at a pressure $< 2 \times 10^{-6}$ Torr. Figure 1(b) contains representative linear and semi-log transfer curves of a single-layer MoS$_2$ FET that shows n-type behavior. The transfer curves were acquired at a bias ramp rate of 10 V/s in steps of 1 V. The field-effect mobility ($\mu_{FE}$) is calculated from these curves according to the following equation:

$$\mu_{FE} = \frac{dI_D}{dV_G}\left[\frac{L}{WC_iV_D}\right]$$

where $I_D$ is the drain current, $V_G$ is the gate voltage, $C_i$ is area-normalized capacitance of 300 nm thick $SiO_2$ (11 nF/cm$^2$), $V_D$ is the drain voltage, and $L$ and $W$ are the length and width of the channel, respectively ($L$ = 4 µm and $W$ = 9.9 µm for the data shown in Figure 1). The field-effect mobility for this device was found to be 65 cm$^2$/Vs with an on/off ratio of 10$^5$. The sub-threshold swing extracted from Figure 1(b) has a value of ~2 V/decade. It should be noted that our values for the field-effect mobility are an underestimate since we do not exclude the contact resistance in our two-probe measurements. An upper estimate on the field-effect mobility values can be obtained by eliminating the series contact resistance (~18 Ω.mm)[27] of Au on MoS$_2$. This correction raises the mobility value by a factor of ~2.7 ± 0.6, which is consistent with recent observations[4,28] (Supplementary Material S4[46]). Further analysis of the field-effect mobility versus gate voltage ($V_G$) shows a power law behavior in the sub-threshold region, which then saturates in the linear region (Supplementary Material S5[46]).

To illustrate the importance of measuring unencapsulated single-layer MoS$_2$ FETs in vacuum, we also performed measurements in ambient conditions. A comparison of the transfer plots (from the same device shown in Figures 1(a) and (b)) in ambient and vacuum is provided in Figure 1(c). The linear field-effect mobility in ambient is reduced by a factor of 5-8 compared to vacuum, thus revealing the deleterious effects of atmospheric adsorbates on charge transport in single-layer MoS$_2$. In addition, the observed shift in threshold voltage suggests that atmospheric adsorbates induce substrate doping and/or trapped charge. Similar effects have also been observed on bilayer MoS$_2$ devices, where adsorbed oxygen and water were implicated in degraded device performance.[29]

To further investigate the charge transport mechanism, variable temperature measurements were performed on the same device at a pressure of < 2 x 10$^{-6}$ Torr. Figure 2(a) shows the resulting threshold voltage subtracted linear transfer curves over the temperature range of 5 to 295 K, where the drain current at constant carrier concentration increases with decreasing temperature. This observation is in contrast to the recent report by Ghatak *et al.*,[18] where Mott variable range hopping (VRH) transport was observed at all $V_G$, although the behavior at higher $V_G$ was not reported. On the other hand, Radisavljevic *et al.*[19] and Ghatak *et al.*[18] found that the drain current diminished considerably with reducing temperatures, leading to conclusions of thermally activated and Mott VRH-like transport, respectively.

Figure 2(b) shows the transfer plots on a semi-log scale. The on-currents ($I_D$ at $V_G$ = 40 V) are nearly unchanged with temperature. Similarly, the off-currents ($I_D$ at $V_G$ < -60 V), which are limited by the noise floor of the measurement apparatus, also appear to remain unchanged, leading to the on/off ratios maintaining the same order of magnitude (~10$^5$) at all measured temperatures. The threshold voltage, however, shifts to more positive values with decreasing temperature (see Supplementary Material S6[46]), which implies a rise in transconductance with reducing temperature as is also seen in Figure 2(a).

In an effort to identify the dominant charge transport mechanism, we attempted fitting our two-probe conductivity data with the Mott VRH model given by:

$$\sigma = \sigma_0(T)\exp\left[-\left(\frac{T_0}{T}\right)^{\frac{1}{d+1}}\right]$$

where $\sigma$ is the measured conductivity, $T$ is the temperature, $d$ is the dimensionality of the system, $\sigma_0$ is a temperature dependent conductivity prefactor given by $AT^m$, where $m$ varies from 0.8 to 1, $A$ is a constant, and $T_0$ is the characteristic temperature given by the slope of the linear fits for σ versus $T^{-1/3}$. We choose $d = 2$ in this case since a stoichiometric monolayer material would maintain charge transport strictly in two dimensions as has been verified previously for ultra-thin $MoS_2$.[18] The VRH model is usually used to describe charge transport in the case of highly disordered systems[30] where the electronic states are discretely localized, instead of forming bands with a continuous density of states. However, even in the case of highly crystalline materials such as $MoS_2$, the presence of a high density of localized states in the band-gap region can lead to hopping transport when the Fermi level passes through them. Figure 2(c) shows the temperature-dependent conductivity data and fits using the above Mott VRH equation for $m = 0.8$. A good fit ($r^2 > 98\%$) for all $V_G$ indicates that the charge transport is dominated by the Mott VRH mechanism (extracted localization lengths are shown in Supplementary Material S7[46]). Similar fits were also attempted using the thermally activated charge transport model. However, our data was found to fit poorly ($r^2 < 75\%$) to the thermal activation model, especially at high $V_G$ ($V_G > 0V$), and hence thermal activation was ruled out.

The rise in field-effect mobility with reducing temperatures as seen in Figure 2(a) has been commonly observed for crystalline inorganic semiconductors,[31-34] graphene,[35] and highly ordered organic small molecule thin films.[36] This increase in transconductance/mobility with reducing temperature has previously been attributed to band-like transport.[36] This mechanism can be further justified by considering the position of Fermi level and occupancy of trap states in the gap region. At low carrier densities ($V_G < V_{th}$), the Fermi level lies in the region of localized trap states inside the gap. Consequently, transport in the sub-threshold regime is expected to be VRH, similar to the disorder-induced localization at low carrier densities previously observed in Si.[37,38] With increasing gate bias, the Fermi level moves closer to the conduction band thus filling up the trap states and entering the band-tail of mobile states in the gap. Alternately, the carriers may also get thermally excited to the conduction band at ($V_G > V_{th}$) which could lead to the observed band-like transport behavior.[38] Since our observation of band-like transport contrasts recent reports in other unencapsulated MoS$_2$ devices,[18,19] it appears that large sample-to-sample variations exist. We also note that adsorbate condensation during cryogenic cool-down led to VRH behavior at the same biasing conditions. In particular, we observed that lower vacuum levels (> 5x10$^{-6}$ Torr) lead to visible condensation on the sample surface, which results in diminishing currents and VRH up to $V_G$ = 50 V ($n$ = 3.6 x 10$^{12}$ cm$^{-2}$). Consequently, to observe the intrinsic MoS$_2$ band-like transport, the devices not only need to be carefully prepared but also measured under conditions that minimize surface adsorbates.

To gain further insight into the charge scattering mechanism, the field-effect mobility of three devices is plotted as a function of temperature in Figure 3. In all cases, the field-effect mobility follows a power law dependence with temperature for T > 100 K given by $\mu_{FE} \alpha\ T^{\gamma}$, where the average value of the exponent $\gamma$ is 0.62±0.13. A power law dependence with a positive exponent is indicative of a phonon scattering mechanism,[39] which is consistent with other materials that show band-like transport such as graphene,[35] ultrathin Bi$_2$Te$_3$,[34] few layer MoSe$_2$,[40] and other crystalline inorganic semiconductors.[31-33,41-44] Recent theoretical calculations for MoS$_2$ estimate the exponent $\gamma$ to be ~1.7. Since no scattering from substrate phonons was considered in this model, it is likely that polar optical phonons from the underlying oxide may also play a role in the overall scattering at these temperatures. The exact value of the exponent $\gamma$ ~ 0.62 in our devices may also be affected by variations in the effective

Schottky barrier height with temperature,[27] although our output curves were found to remain linear in the reported temperature range (Supplementary Material S8[46]). Below 100 K, the field-effect mobility saturates to an intrinsic value limited by Coulomb scattering as has also been observed in the case of unencapsulated single-layer graphene FETs.[35]

In conclusion, we have fabricated and characterized unencapsulated single-layer $MoS_2$ FETs with room temperature field-effect mobilities in excess of 60 cm$^2$/Vs. Although these field-effect mobility values are significantly higher than previous reports on unencapsulated devices, they remain lower than theoretically predicted values of ~400 cm$^2$/Vs.[39] Since $MoS_2$ FETs were found to be sensitive to extrinsic adsorbates, the substrate/dielectric is likely to become the limiting factor in the presence of suitable encapsulation methods. Alternative substrates will then be necessary to realize mobility values closer to the theoretical limit. One such example of substrate-induced mobility enhancement is the case of graphene on hexagonal boron nitride.[45] Additionally, the crystalline quality of the sample could also be contributing to the lower observed mobility values compared to theoretical predictions. We further conclude that the charge transport in unencapsulated $MoS_2$ FETs is dominated by Mott variable range hopping in the sub-threshold regime and band-like transport in the linear regime, assuming that appropriate measures are taken to minimize extrinsic disorder including contamination and condensation. Overall, these results highlight the critical parameters that underlie charge transport in $MoS_2$ and thus will help guide future efforts to realize high performance $MoS_2$-based electronic and optoelectronic applications.

**Acknowledgments**

This research was supported by the Materials Research Science and Engineering Center (MRSEC) of Northwestern University (NSF DMR-1121262). The authors thank B. Myers for assistance with electron beam lithography. D.J.L. would like to thank the Indo-US Science & Technology Forum (IUSSTF) for a postdoctoral fellowship and Prof. C.N.R. Rao for helpful discussions. J.E.J acknowledges an IIN Postdoctoral Fellowship provided by the Northwestern University International Institute for Nanotechnology. This research made use of the NUANCE Center at Northwestern University, which is supported by NSF-NSEC, NSF-MRSEC, Keck Foundation, and the State of Illinois.

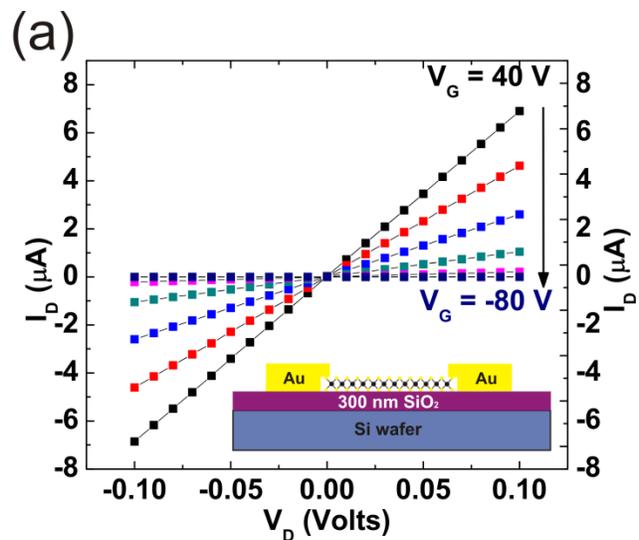

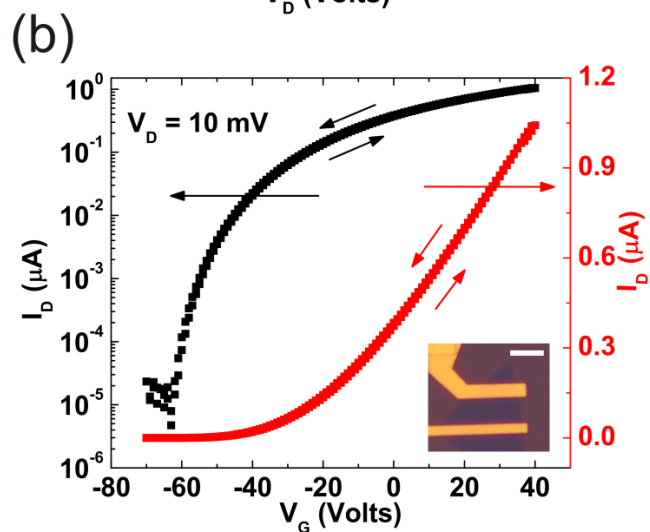

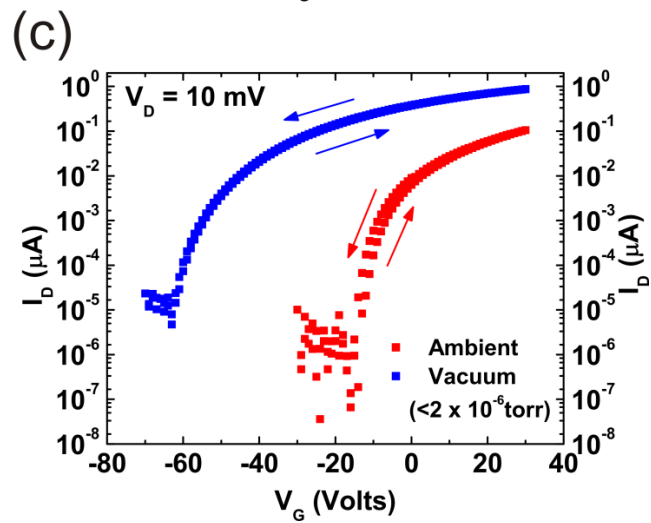

**Figure 1.** Field-effect transistor (FET) characteristics: (a) Output plots of a single-layer $MoS_2$ FET showing linear I-V characteristics indicative of electrically ohmic contacts. Inset shows the schematic of the device. (b) Linear and semi-log transfer plots of a representative single-layer $MoS_2$ FET. The shorter arrows indicate the sweep direction. The longer arrows indicate the appropriate y axis. The inset shows an optical micrograph of the device. The scale bar is 5 µm. (c) Comparison of transfer plots of the same device under ambient (red) and vacuum (< $2 \times 10^{-6}$ Torr) (blue).

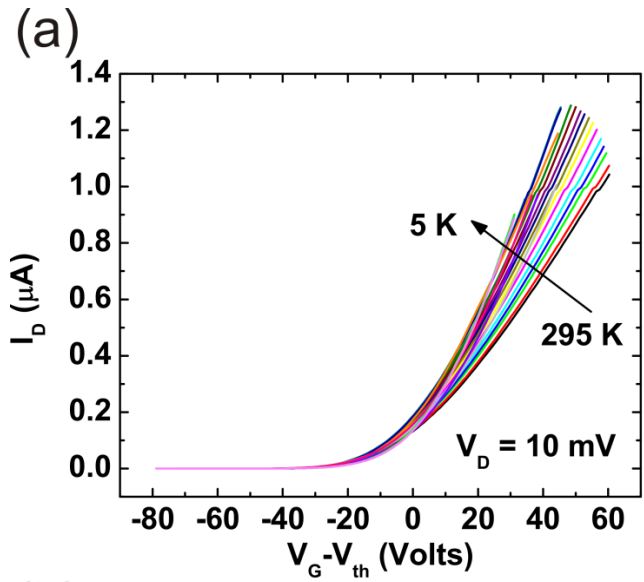

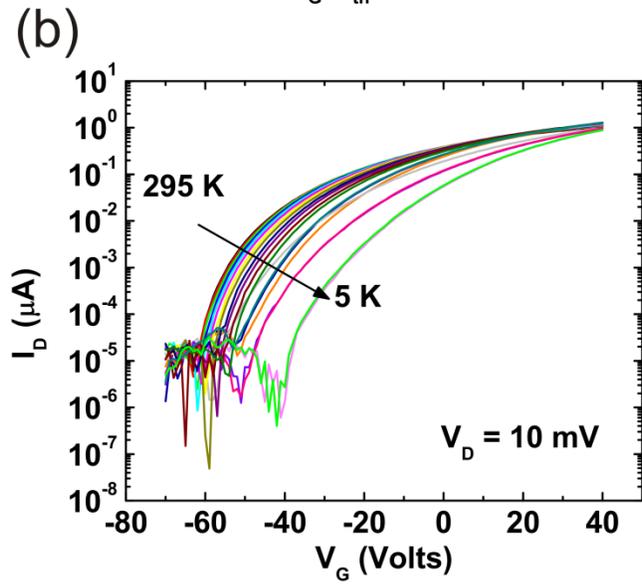

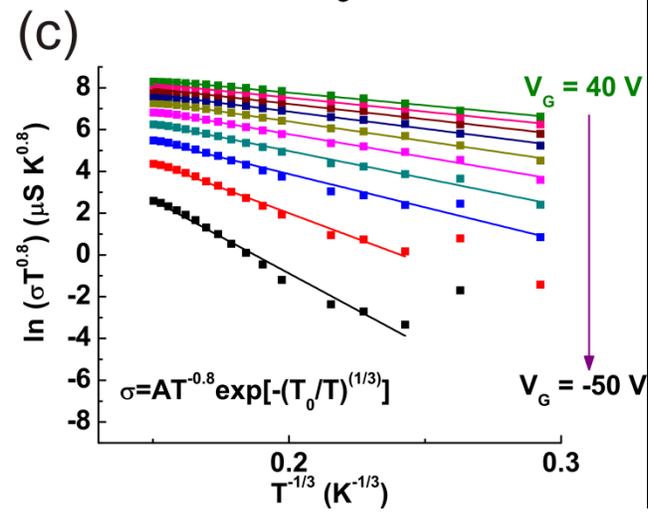

**Figure 2.** Variable temperature transport: (a) Threshold voltage normalized linear transfer plots for temperatures between 5 and 295 K. (b) Semi-log transfer plots for temperatures between 5 and 295 K. (c) Two probe conductivity fits to the Mott variable range hopping (VRH) model for a range of gate biases.

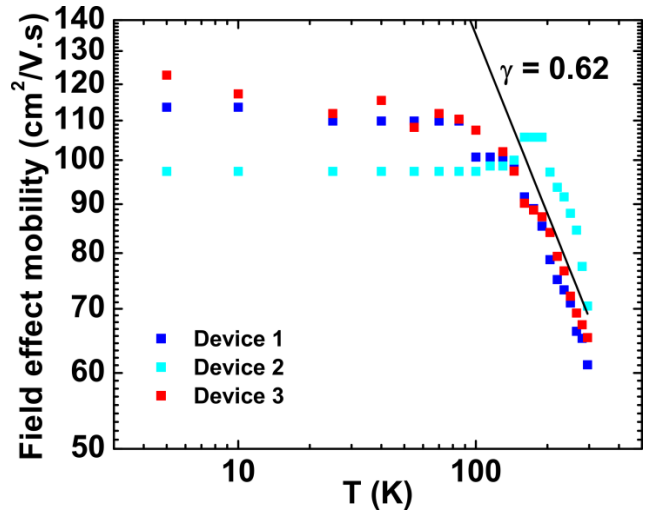

**Figure 3.** Temperature-dependent field-effect mobility for three single-layer MoS$_2$ FETs. At higher temperatures (T > 100K), the field-effect mobility ($\mu_{FE}$) follows $\mu \propto T^{-\gamma}$ with $\gamma$ = 0.62±0.13 before saturating to a constant value at low temperatures (T < 100 K). The black line ($\gamma$ = 0.62) is included to guide the eye.